# A simplified model based on self-organized criticality framework for the seismic assessment of urban areas

A. Greco[a*], A. Pluchino[b], L. Barbarossa[a], I. Caliò[a], F. Martinico[a], A. Rapisarda[b,c]

[a]*Department of Civil Engineering and Architecture,University of Catania, viale A. Doria 6, Catania, Italy*

[b]*Department of Physics and Astronomy, University of Catania, and INFN Sezione di Catania, via S.Sofia 64, Catania, Italy*

[c] *Complexity Science Hub, Vienna, Austria*

**Abstract**

The analysis of the seismic vulnerability of urban centres has received a great attention in the last century, due to the progressive concentration of buildings in metropolitan areas. In order to estimate the seismic vulnerability of a densely populated urban area, it would in principle be necessary to develop in-depth analyses for predicting the dynamic behaviour of the individual buildings and their structural aggregation. Such analyses, however, are extremely cost-intensive, require great processing time and above all expertise judgement. It is therefore very useful to define simplified rules for estimating the seismic vulnerability of whole urban areas by simulating different types and intensities of seismic stresses. In the last decades, the Self-Organized Criticality (SOC) scenario has gained increasing credibility as a mathematical framework for explaining a large number of naturally occurring extreme events, from avalanches to earthquakes dynamics, from bubbles and crises in financial markets to the extinction of species in the evolution or the behaviour of human brain activity. All these examples show the intrinsic tendency common to many phenomena to spontaneously organize into a dynamical critical state, whose signature is the presence of a power law behaviour in the frequency distribution of events. In this context, the Olami-Feder-Christensen (OFC) model, introduced in 1992, has played a key role in modelling earthquakes phenomenology. The aim of the present paper is proposing an agent-based model of earthquake dynamics, based on the OFC self-organized criticality framework, in order to evaluate the effects of a critical sequence of seismic events on a given large urban area during a given interval of time. The further integration of a GIS database within a software environment for agent-based simulations, will allow to perform a preliminary parametric study of these effects on real datasets. The model could be useful for defining planning strategies for seismic risk reduction.

\* Corresponding author. Tel.: 00390957382251;
*E-mail address:* agreco@dica.unict.it

## 1. Introduction

In many cases, large seismic events are not isolated but are preceded and/or followed by a foreshock-aftershock activity of variable intensity and duration. For example, the severe earthquake of magnitude 5.9 ML occurred in L'Aquila (Italy) on April 6 2009, at 3:32 a.m., that caused more than 300 victims, 1,600 wounded and more than 10 billion euros of estimated damages, was the mainshock of an anomalous activity which started in December 2008 and lasted until 2012. In order to give an idea of the great number of shocks involved it is interesting to highlight that just in the year that followed the April 6 event, the Italian institute for geophysics and volcanology (INGV) reported that about 18,000 earthquakes occurred only across the area of the city of L'Aquila with different epicenters (256 events were registered only during the 48 hours immediately after the mainshock, 56 of them with a magnitude greater than 3 ML).

It is quite natural to frame this kind of phenomenon in the context of the Self-Organized Criticality (SOC) paradigm. Introduced in 1987 by Bak, Tang and Wiesenfeld [1], SOC theory states that many large interactive systems observed in nature can self-organize into a "critical state". Once in this state, small perturbations may result in chain reactions, which can affect any number of elements within the system. A commonly used, intuitive example calls upon the hypothetical generation of a pile of sand. Imagine sand being added one grain at a time to a sandbox. At first, the grains land harmlessly on the stable slope of a proto-sand pile. As more grains are added the slope of the pile increases. Eventually, the slope locally reaches a critical value such that the addition of even one more grain may result in an "avalanche". The key point is that the next avalanche can be of any size, ranging from a single grain to a catastrophic collapse of the sand pile. Moreover, the size distribution of the avalanches follows a characteristic power law which indicates a critical behaviour that compare well with experimental data.

Since Bak's original paper, a lot of phenomena of strikingly different backgrounds were claimed to exhibit SOC behavior: sandpiles, earthquakes, forest fires, rivers, mountains, cities, literary texts, electric break-down, motion of magnetic flux lines in superconductors, water droplets on surfaces, dynamics of magnetic domains, growing surfaces, human brains, etc. [2] In particular, the dissipative Olami-Feder-Christensen (OFC) model adopts the SOC hypothesis in order to reproduce the scale-invariant dynamics of real earthquakes on a regular square lattice, which mimics a portion of terrestrial crust [3]. When, after a given transient, the system enters into a critical state, the average earthquakes activity increases and events of any scale *may* occur. This is probably what happened between 2008 and 2012 in the territory of L'Aquila: the region entered into a critical state, and at that point the probability to experience a large earthquake, like that one of April 6 2009, was no more negligible, even if – as a consequence of the SOC dynamics – it would have been impossible to predict the exact moment in which that event would have been realized.

In this paper, by adopting the SOC framework of the OFC model, we tried to reproduce a situation similar to that one observed in L'Aquila region, but choosing as a case study the territory around Avola, a small city in the southeast part of Sicily, a zone which, from the point of view of the maximum observed macro-seismic intensity, is very similar to the area around L'Aquila. The aim of the study is to test, through an agent-based simulative approach, the seismic vulnerability of that urban area, based on real features of existing buildings, under the assumption that the crust below it would be set up into a critical state, i.e. that would experience a long sequence of earthquakes of any size with epicenters located in different part of the considered territory.

The paper is organized as follows. In Section 2, simple rules for estimating the seismic vulnerability of buildings in urban areas are described. In Section 3, the dynamics of the OFC model, and its adaptation to the classification adopted in the European Macroseismic Scale [4], are presented. In Section 4, the case study of Avola and the GIS dataset used in the paper are discussed. Finally, in Section 5, the results of the numerical simulations are shown, then some conclusions are drawn.

## 2. Simplified seismic vulnerability estimate of existing buildings in urban areas

The estimate of the seismic vulnerability of existing buildings has been extensively studied during the last 30 years by using different levels of approximation. In fact a reliable vulnerability evaluation for a single building requires expert analytical calculations and a deep knowledge of the geometry of the structure and of its mechanical properties. These kinds of analyses are therefore unsuitable to perform a vulnerability assessment at urban scale, which must be based on simplified approaches and rapid processing. For this reason several procedures for a synthetic assessment of the seismic vulnerability of either masonry or reinforced concrete existing buildings have been presented in the scientific literature [5-9] and also adopted in national and regional codes [10,11].

The grade of vulnerability of a structure must of course be related to the possible seismic actions on the specific site, which can only be statistically presumed from previous recorded data. It is important to highlight that not only severe ground motions constitute a danger for structures since damage can occur even for moderate actions and therefore a building can collapse after several small earthquakes due to incremental damage.

In a seismic impact evaluation at regional or urban scale it would be very useful to have the possibility to estimate both the collapse scenario under severe earthquakes and the progressive one caused by moderate ground shakings.

The fundamental steps, proposed in this paper, to perform subsequent vulnerability analyses at urban scale are summarized in the following:

1) Assessing an initial synthetic vulnerability value for each building by means of the available data.
2) Modeling the seismic ground motions in the considered urban area.
3) Assuming a correlation between the damage produced in each building and the seismic intensity.
4) Evaluating the global seismic response of the considered area in terms of damage parameters.
5) Updating the seismic vulnerability of each structure taking into account its eventual damage
6) Starting again from step two for a new seismic input.

Either in the first step of the previously described analysis or in the successive ones it is important to dispose of an estimate of the vulnerability of the structure. The evaluation of the synthetic vulnerability value for each building must take into account several parameters (which are different for masonry and reinforced concrete buildings) which among others consider the geometry of the structure, the mechanical properties of the material, the quality of the construction and the geological characteristics of the site. For example in the DPC-DRPC data sheet of the Regional Civil Protection Department in Sicily [11] the following 7 different parameters are considered for reinforced concrete buildings: age, quality of the resistant system, average normal tension of the 1st level columns, regularity in plan, type of infill at the 1st level, presence of non-structural elements that may collapse causing damage, location of the building and foundations.

Masonry buildings are instead analyzed by means of the following 9 parameters: efficiency of the connections, quality of the resistant system, location of the building and foundations, constructive masonry resistance, quality of horizontal structures, planimetric configuration, resistance of coverage structures, presence of non-structural elements that may collapse causing damage, presence of damage.

For both the typologies, in relation to the value assumed by each single indicator, a score is associated which represents a vulnerability index V for the building.

A reliable estimate of the seismic vulnerability of an existing building needs therefore a significant amount of data even for a synthetic appraisal.

In absence of sufficient information on the building, a representative vulnerability index can anyway be assumed following some approximate approaches presented in the scientific literature. In this paper as proposed in [5], the vulnerability index V will be chosen in a fixed range (conventionally defined from −0.02 to 1.02) characteristic of building typologies deduced from observed damage data, as reported in the following table:

Table 1

| Typologies | Building type | Vmin | Vmax |
|---|---|---|---|
| Masonry | M1 Rubble stone | 0.62 | 1.02 |
| | M2 Adobe (earth bricks) | 0.62 | 1.02 |
| | M3 Simple stone | 0.46 | 1.02 |
| | M4 Massive stone | 0.3 | 0.86 |
| | M5 U Masonry (old bricks) | 0.46 | 1.02 |
| | M6 U Masonry—r.c. floors | 0.3 | 0.86 |
| | M7 Reinforced /confined masonry | 0.14 | 0.7 |
| Reinforced Concrete | RC1 Frame in r.c. (without E.R.D) | 0.3 | 1.02 |
| | Frame in r.c. (moderate E.R.D.) | 0.14 | 0.86 |
| | Frame in r.c. (high E.R.D.) | -0.02 | 0.7 |
| | RC2 Shear walls (without E.R.D) | 0.3 | 0.86 |
| | Shear walls (moderate E.R.D.) | 0.14 | 0.7 |
| | Shear walls (high E.R.D.) | -0.02 | 0.54 |

For the intensity of the seismic input the classifications used in the European Macroseismic Scale (EMS) with the following 12 levels is adopted: I. Not felt, II. Scarcely felt, III. Weak, IV. Largely observed, V. Strong, VI. Slightly damaging, VII. Damaging, VIII. Heavily damaging, IX. Destructive, X. Very destructive, XI. Devastating, XII. Completely devastating. This classification can be related to the Richter Magnitude scale whose maximum intensity is set to 8.5. For example levels III, VI and VIII in the EMS correspond respectively to intensities in the Richter scale of about 2.5, 4.3, 5.5.

In the present study the seismic intensity is considered as a continuous parameter in the range 1-12 evaluated with respect to a rigid soil condition; possible amplification effects due to different soil conditions are accounted for inside the vulnerability parameter V.

The correlation between the seismic input and the expected damage $\mu_D$, as a function of the assessed vulnerability, is expressed in terms of vulnerability curves described by a closed analytical function [5]:

$$\mu_D = 2.5 \left[ 1 + \tanh\left( \frac{I + 6.25V - 13.1}{Q} \right) \right] \tag{1}$$

where I is the seismic input provided in terms of a macroseimic intensity, and V and Q are, respectively, the vulnerability and the ductility index. For the ductility index, the value Q = 2.3 has been assumed, judged to be representative for buildings not specifically designed to have ductile behavior. When a r.c. building is designed according to seismic code prescriptions its ductility index is assumed to be Q = 2.6 . It is worth to point out that increasing the value of Q, flattened curves are obtained, representative of more ductile behavior, as less damage increase is observed for the same increase in seismic input.

According to the EMS-98 [4], 5 damage grades have been considered,:
$\mu_D = 1$ slight, $\mu_D = 2$ moderate, $\mu_D = 3$ heavy, $\mu_D = 4$ very heavy, $\mu_D = 5$ destruction, plus the absence of damage $\mu_D = 0$ no damage.

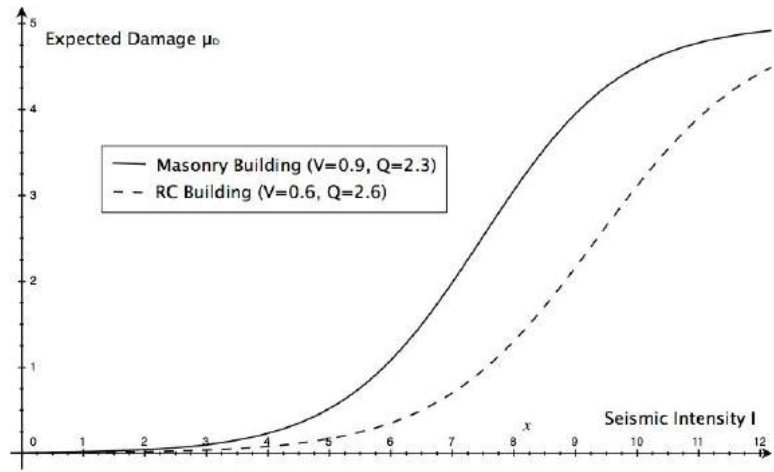

Figure 1 Example of expected damage vs seismic input

The evaluation of the expected damage for each building allows to globally visualize at the urban scale the areas with the same level of damage after each seismic input. Damage will successively proportionally modify the vulnerability of each structure which can therefore became progressively inadequate to stand successive ground motions.

The various parameters, and the procedure steps, introduced in this section will be reconsidered and further specified afterwards. Before this, more details are needed about the OFC model of earthquakes and the GIS dataset adopted in this paper.

### 3. OFC: a self-organized criticality model of earthquakes

In recent years there has been an intense debate on earthquake predictability and a great effort in studying earthquake triggering and interaction. Along these lines the possible application of the self-organized criticality (SOC) paradigm has been discussed. Earthquakes trigger dynamic and static stress changes. The first acts at short time and spatial scales, involving the brittle upper crust, while the second involves relaxation processes in the asthenosphere and acts at long time and spatial scales. In [12] it has been shown that it is possible to reproduce statistical features of several earthquakes catalogues within a SOC scenario taking into account long-range interactions in the context of the dissipative Olami-Feder-Christensen model implemented on a *small world* topology.

The Olami-Feder-Christensen (OFC) model [3] is one of the most interesting models displaying self-organized criticality. Despite its simplicity, it exhibits a rich phenomenology resembling real seismicity, such as the presence of aftershocks and foreshocks. In its original version the OFC model consists of a two-dimensional square lattice of $N = L^2$ sites, each one connected to its four nearest neighbours and carrying a seismogenic force (seismic stress) represented by a real variable $F_i$, which initially takes a random value in the interval [0 , $F_{th}$]. In order to mimic a uniform tectonic loading all the forces are increased simultaneously and uniformly, until one of them, say the *i*-th, reaches the threshold value $F_{th}$ and becomes "active" ($F_i \geq F_{th}$, where typically, $F_{th}$=1). The driving is then stopped and an "earthquake" or avalanche can start: actually, the active site transfers all its stress on its nearest-neighbors, each of which receives a fraction α of seismic force and, in turn, can overcome its own threshold becoming active, and so on and so forth. The dynamical rule is the following:

$$F_i \geq F_{th} \rightarrow \begin{cases} F_i \rightarrow 0 \\ F_{nn} \rightarrow F_{nn} + \alpha F_i \end{cases} \tag{2}$$

where "nn" denotes the set of nearest-neighbour sites of i. The number of topplings (active sites) during an avalanche defines its size S, while the dissipation level of the dynamics is controlled by the parameter α. The model is conservative if α = 0.25, while it is dissipative for α < 0.25. In this paper, a dissipative version of the OFC model, with α = 0.21, is considered and it is implemented on a regular lattice network with N = 1600 nodes (L = 40) and open boundary conditions, i.e., F = 0 on the boundary nodes. In order to improve the model in a more realistic way, a small fraction of long-range links in the network has been introduced, in order to obtain a small world network topology. Just a few long-range edges create shortcuts that connect sites which otherwise would be much further apart. This kind of structure allows the system to synchronize and to show both finite-size scaling and universal exponents. Furthermore, a small world topology is expected to model more accurately earthquakes spatial correlations, taking into account long-range as well as short-range seismic effects.

In the version of the OFC model here adopted, the links of the regular lattice are rewired at random with a probability p=0.02, a value that allows to obtain both small world features and criticality. The resulting network is shown in Figure 2(a), where the brightness of each node in gray-scale color is proportional to its level of seismic stress. In the top panel of Figure 2(b) the size S of about 3000 subsequent earthquakes during a typical run of the OFC dynamics is plotted. After a transient of about 1500 events, where the maximum size $S_{max}$ involves less than 5% of the entire lattice, the system enters into a critical state, where the average size of the avalanches starts to increase and large events, involving a great number of nodes, have a non-zero probability of occurrence. In the reported sequence, about 10 events in the critical state showed a size greater than N/3, with a maximum $S_{max} \approx$ N/2. The nodes activated during one of these big events are colored in red in Figure 2(a). The presence of criticality in the earthquake sequence is revealed by a power-law distribution (pdf) of the avalanches' size, which appears as a straight line in the log-log plot shown in the bottom panel of Figure 2(b). The power-law is also the signature of a scale-invariant behaviour of the shocks, meaning that the size distribution of the avalanches appears the same at each spatial scale.

In order to adapt the OFC model output to the classification used in the European Macroseismic Scale (EMS), one needs to transform the size S of a given earthquake into an intensity *I*, which – as already said – presents 12 different possible levels. The transformation function is the following: $I = 12\beta S/N$ , where β is a tuning parameter whose role is to ensure that the maximum size would not exceed the maximum intensity level. Usually, a value β=1.8 fits well with this purpose (an earthquake size S≈900 would correspond to an intensity I≈12). In the next section, the case study of Avola is introduced. Then, the integration of the OFC model with the Avola Gis database will be discussed, together with the first simulation results.

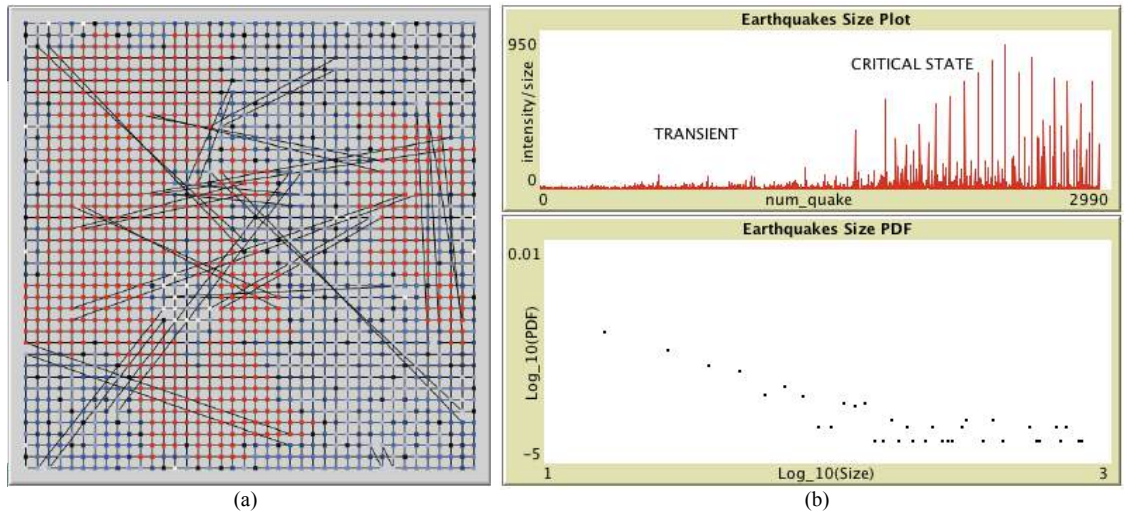

Figure 2  (a) The small-world lattice of the OFC model (active nodes in red); (b) The sequence of earthquakes' sizes in the
transient and in the critical state (upper panel) with the corresponding power-law distribution (lower panel).

### 4. The case study of Avola

The city of Avola (31576 inhabitants in 2016) is located along the south-east coast of Sicily, the so-called Val di Noto, thirty kilometers south of Syracuse. It was completely destroyed in 1693 by a major earthquake that hit South-eastern Sicily, causing thousand of victims. More than 45 cities were destroyed or severely damaged. This catastrophic event caused a complete change in the structure of the entire Val di Noto area, where a number of cities were rebuilt in new sites, closer to the coast.

After the earthquake, also the city of Avola was rebuilt in a new site according to a completely new layout in the coastal plain, 1 kilometers far from the coastline. The urban structure is characterized by a grid of perpendicular streets within an hexagonal perimeter. A large main square with nearby minor ones marks the heart of town, according to a design inspired by the ideal cities plans from the Renaissance. Until the end of 19th century, the urban growth around the early urban core was influenced by the hexagonal shape of the settlement (Fig. 3), made by concentric blocks, somewhere irregular. The pattern based on compact and regular rectangular blocks repeats the model of the agro towns founded in Sicily from 15th to 17th century, especially during the Spanish domination. At the beginning, the regular grids were aligned to the sides of the hexagon and after their layout was oriented by the grid of existing long distance and rural roads. This phase addressed the relevant demand of urban growth between the 1940 and 1960.

Between 1970 and 1990, along with the urban core development, two new processes molded the shape of the settlement. The first was the development of extensive subdivisions with detached single family holiday houses along the coastline, a phenomenon that overwhelmed the fragile coastal ecosystem, the second one was the low-density urbanization of peri-urban and rural areas where a considerable number of small and medium size houses have been built by the land owners for week-end or seasonal usage.

Recently the, urban growth processes have been governed by poor quality urban plans that gives marginal attention to agricultural land protection and sustainability. The result are the new medium density settlements, developed close to the town center, following an awkward interpretation of the modernist planning models [13].

## 4.1. Description of the GIS dataset

During the new city masterplan design process, carried out from 2013 to 2016, data were collected, digitized and georeferenced, to analyze urban growth of the city. The study was based on all the historical cartographies available, it produced a map representing the growth of Avola settlement from the foundation in the early 18th century, from 2015.

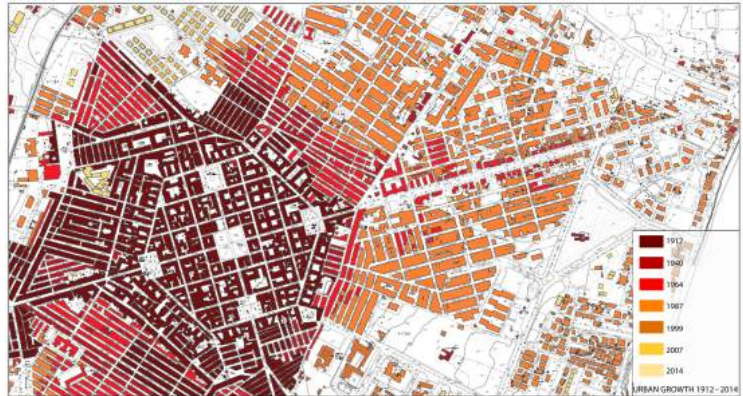

Figure 3 City of Avola - Urban Growth map

Historical cartographies of urban fabric were overlaid with new official cartography, released by Urban Planning Department of the Regional Government, in order to obtain an historical dating of the entire built up area. As a result, urban growth had been quantified and mapped measuring the built-up changes corresponding to seven dates (1912, 1940, 1964, 1987, 1999, 2007, 2015).

The resulting urban growth map gives for each building of the urban fabric the date in which it is present in the corresponding map. This allows an estimate of the period of construction for each building. In addition, using the data (height and surface) derived from the official vectorial cartography, the volume of each building of the urban fabric has been computed by using standard Gis functions. As a result, every building in Avola had been characterized by its volume, height and construction date attributes in the Gis dataset. In addition, the dataset includes the same information for other buildings scattered in the territory around Avola. In particular, a square area with a side length of 10,5 Km has been considered, as shown in Figure 4(a). The total number of buildings in this area is $N_B$=25830. Depending on their period of construction, all the buildings were classified in two main categories, reported in Figure 4(b) with different colors: masonry buildings (before 1965, in brown) and reinforced concrete buildings (after 1965, in gray). Buildings of the first category will present a ductility Q=2.3, while those of the second category will have Q=2.6. Then, crossing the construction information with data about the ratio R=H/L between height H and base side L, a vulnerability index V has been assigned to each building following the prescriptions of Table 1: in Figure 4(c) we represent in green buildings with low vulnerability (-0.02<V<0.3), in yellow those with medium vulnerability (0.3<V<0.65) and in red those with high vulnerability (0.65<V<1.02). Notice that most of the highly vulnerable buildings coincides with the masonry ones and are concentrated in the urban area of Avola, while many of the buildings spread in the peri-urban territory are less vulnerable, since they are mainly made of reinforced concrete.

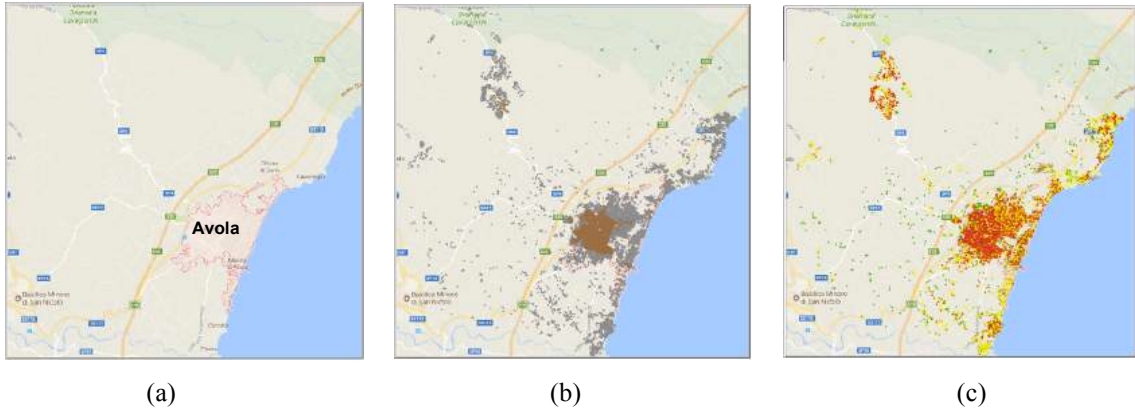

(a)                                   (b)                                   (c)

Figure 4 (a) Territory of Avola; (b) Gis dataset: Masonry (brown) and reinforced concrete (gray) buildings; (c) Low (green), medium (yellow) and high (red) vulnerability buildings.

## 5. Discussion and Simulation results

In this section, the OFC model introduced in section 3 has been integrated with the Gis dataset presented in the previous section in order to evaluate, through an agent-based simulation, the impact of a sequence of earthquakes on the vulnerability of the buildings present in the territory of Avola. The idea is to simulate a period of several months during which the considered area, like the L'Aquila territory in 2008-2009, is supposed to be in a critical state. Therefore, thousands of shocks of any size, and with different epicenters in the same area, will invest the buildings, by producing damages which depends on their ductility and vulnerability. Every damage suffered can progressively enhance the vulnerability of the buildings, until a given strong shock will be able to destroy some of them. At the end of the simulation, the fraction of undamaged, damaged and destroyed buildings over the total $N_B$ will be a good indicator of the global response of the territory to the seismic inputs.

In Figure 5 the OFC small-world lattice with N=1600 nodes, already shown in Figure 2(a), is reported over the satellite map of the square area around Avola. The distance R between two rows or two columns of the lattice corresponds to about 250 meters on the map. Running the OFC model, after a transient of 1500 small events with intensity less than 2, the system enters in the critical state. Then, a sequence of $N_S$ shocks with any size occurs. As explained in section 3, the size S can be translated in an intensity scale where $I=12\beta S/N$ (with $\beta=1.8$) going between 0 and 12. During an earthquake of size S, each one the S active nodes (colored in red in Figure 5) of the network transfers a seismic stress of the corresponding intensity $I$ to all the buildings around it within a circle of radius R. At this point, according to equation (1), the resulting damage $\mu_D$ for these buildings is evaluated as function of their vulnerability and ductility. Then, each damage further enhances the vulnerability V according to the following rule:

$$V_{new} = V_{old} + \gamma \, \mu_D \, [1 / (1.05 - V_{old})] \qquad (3)$$

where $\gamma$ is a parameter which tunes the effects of the shocks on the buildings. Therefore, subsequent earthquakes can progressively injure undamaged buildings, changing their status in "damaged" when the vulnerability $V_{new} > V_0 + (1.02 - V_0)/2$ (being $V_0$ the initial vulnerability), then in "destroyed" when the vulnerability $V_{new} > 1.02$. Depending on both the maximum intensity $I_{max}$ of the shocks' sequence in the critical state and the value of the parameter $\gamma$, several damage scenarios can be considered.

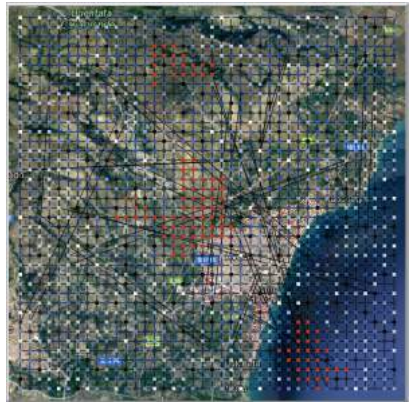

Figure 5 The small world network of the OFC model is superimposed over the satellite map of the Avola's area.

In Figure 6, the simulation results for three different sequences of earthquakes (after the transient), each one with a different value of parameter γ, are shown. For each scenario, the time behavior of the number of both damaged and destroyed buildings is reported on the left, below the earthquakes' intensity time series, while the corresponding map of the considered area, with differently colored buildings (damaged in yellow, destroyed in red), is reported on the right (only nodes of the OFC lattice are visible; all the links have been hidden for a better visualization).

Scenario 1 - In Figure 6 (a), a sequence of about 1000 earthquakes has been considered in the critical state, with a value of $\gamma = 1 \times 10^{-5}$. This value is probably too small to be realistic, since no damage or even any destroyed building is observed until the earthquake intensity remains below 10: only after the last two shocks, with intensities of, respectively, 11 and 12, the number of destroyed buildings (red line) suddenly increases up to 3260 (i.e. 12.6% of the total), which is still very small if compared with the effects expected on the basis of the European Macroseismic Scale (see section 2.2). The red points in the map on the right indicate the location of the destroyed buildings (there are not yellow points, since no buildings were simply damaged).

Scenario 2 - In Figure 6 (b), another sequence of about 2000 earthquakes has been considered, this time with a value of $\gamma = 5 \times 10^{-4}$. In this simulation, the critical state is characterized by a greater number of violent shocks, but with a maximum intensity $I_{max} \approx 9$: as a consequence, the number of damaged buildings (yellow line) starts to increase quite soon, reaching a value of about 1900 at the end of the period considered; at the same time, the number of destroyed buildings (red line) also increases, step by step, in correspondence of each earthquake with $I > 6$, up to a value slightly greater than 5000. The effects of this scenario seem more realistic than those of Scenario 1 and also more in agreement with the EMS predictions. Red and yellow points are clearly visible on the corresponding map on the right.

Scenario 3 - In Figure 6 (c), a last sequence of about 2000 earthquakes has been considered, with a higher value of $\gamma = 1 \times 10^{-3}$. The first 2/3 of the sequence is now characterized by shocks with $I < 7$ which produce a progressive, slight increment, of both the number of damaged and destroyed buildings; then, suddenly, the average intensity of the earthquakes increases and the last seven of them, with an intensity $10 < I < 12$, produce a steep jump in the number of destroyed buildings, which rapidly goes beyond 10000, reaching a final value of more than 15000; simultaneously, in correspondence of the this jump, the number of damaged buildings goes to zero, since all of them have been destroyed by the first mainshock. In the map on the right, the red points have, now, invaded both the urban and the peri-urban area. Also this Scenario seems quite realistic, and in a good agreement with the EMS predictions.

Summarizing, from the comparison with the European Macroseismic Scale, it seems possible to restrict the range of variation of γ within a range between $10^{-4}$ and $10^{-3}$.

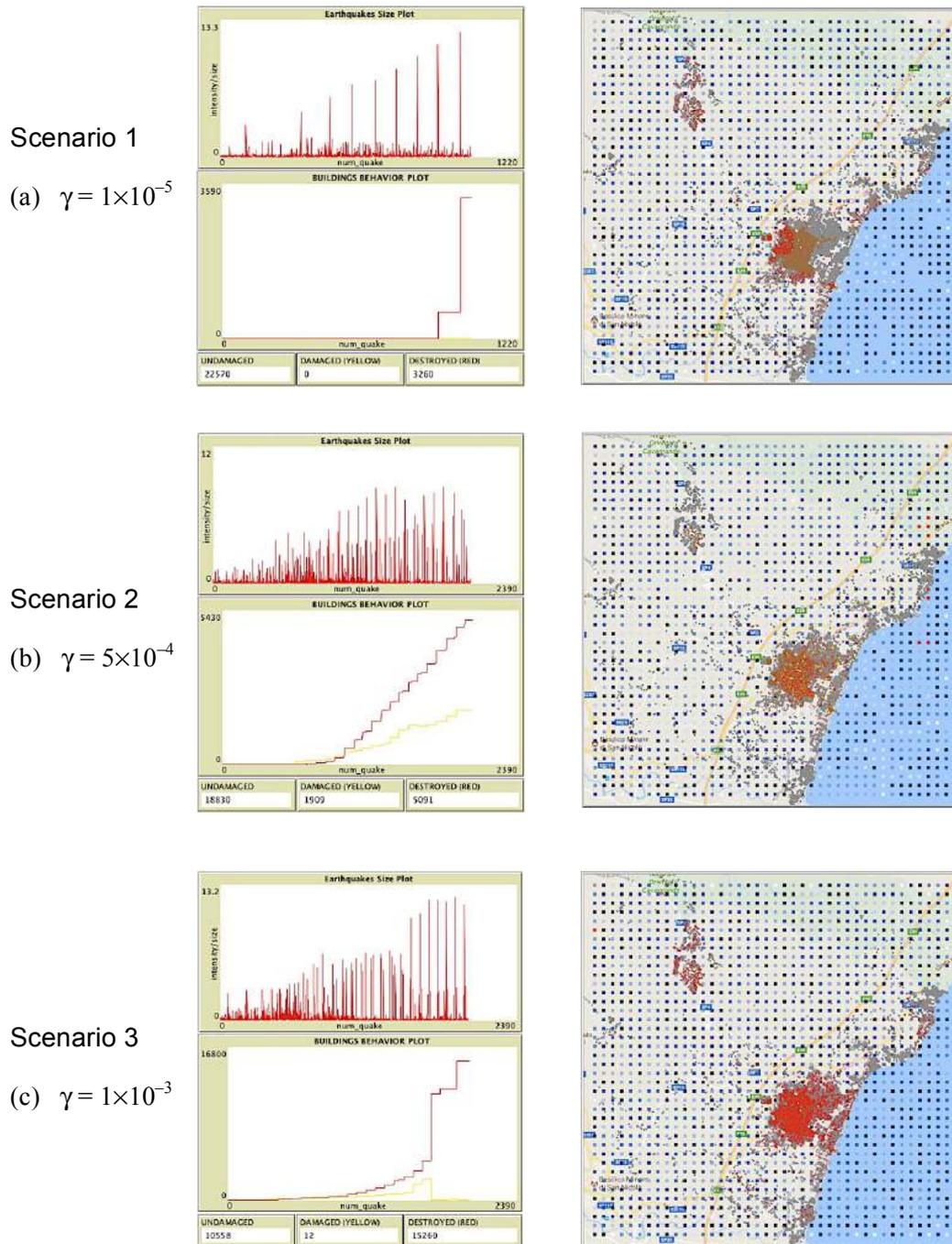

Scenario 1

(a) $\gamma = 1 \times 10^{-5}$

Scenario 2

(b) $\gamma = 5 \times 10^{-4}$

Scenario 3

(c) $\gamma = 1 \times 10^{-3}$

Figure 6  Three different scenarios with different sequences of shocks (after a transient, not reported) and with increasing values of parameter $\gamma$, which tunes the effects of an earthquake on the buildings placed around the active nodes of the lattice.

## 6. Conclusions

This simulative study represents a first attempt to apply a new multidisciplinary agent-based approach to the seismic assessment of an urban, and peri-urban, area. By integrating competences coming from several scientific disciplines, going from the SOC dynamics of earthquakes to the seismic response of buildings with a given vulnerability, from the Gis features of the urban settlement to the agent-based simulations, the proposed methodology allowed to evaluate the effects of a long sequence of shocks with realistic power-law distributed intensities on the buildings present in the area under investigation. Three different seismic scenarios have been considered, and the numerical results clearly showed the potentialities of the present approach. Of course this is only a preliminary study and a further, more accurate, analysis is in progress. In particular, an extended parametric study to explore the seismic response of buildings as function of the OFC grid pitch, of the geological features of soil and of the fault proximity, of the tuning parameter γ, of the intensity and frequency of the mainshocks, will be performed. The possible role of the interaction between adjacent buildings, with consequent cascading effect, will be also investigated. Furthermore, in the application here reported the proposed agent-based strategy has been applied to the scale of a small town for investigating the distribution of damage at the scale of the single building. A similar approach can be also extended to a greater level by investigating the seismic vulnerability of several homogeneous urban areas interested by common seismo-genetic sources. This could be the case of the Oriental Sicily, whose seismic risk is mainly associated to the Ibleo-Maltese system of faults that were responsible of the great devastating 1963 earthquake.

## References

[1] Bak P, Tang C and Wiesenfeld K. Self-organized criticality: An explanation of the 1/f noise. *Phys. Rev. Lett.* **59**, 381, 1987

[2] Jensen HJ. Self-Organized Criticality: Emergent Complex Behavior in Physical and Biological Systems. Cambridge Lecture Notes in Physics, 1998.

[3] Olami Z, Feder HJS, Christensen K. Self-organized criticality in a continuous, nonconservative cellular automaton modeling earthquakes. *Phys Rev E* 1992; **68**(8):1244–1247

[4] EMS. European Macroseismic Scale, 1998. Conseil de l'Europe. European Seism. Commission. 8 LUXEMBOURG 1998.

[5] Lagomarsino S, Giovinazzi S. Macroseismic and mechanical models for the vulnerability and damage assessment of current buildings. *Bull Earthquake Eng* 2006; **4**:415–443.

[6] Zuccaro G, Cacace F. Seismic vulnerability assessment based on typological characteristics.The first level procedure "SAVE". *Soil Dynamics and Earthquake Engineering* 2015; **69**:262–269.

[7] Benedetti D, Petrini V. On seismic vulnerability of masonry buildings: proposal of an evaluation procedure. *L'industria delle Costruzioni* 1984; **18**:66–78.

[8] Di Pasquale G, Orsini G, A Pugliese, Romeo RW. Damage scenario for future earthquakes. In: proceedings of the 11th European conference on earthquake engineering, Paris; 1998.

[9] Masi A. Seismic vulnerability assessment of gravity load designed R/C frames. *Bull Earthquake Eng* 2003; **1(3)**:371–395.

[10] NTC2008 Consiglio Superiore dei lavori pubblici della Repubblica Italiana, Gazzetta Ufficiale n°29 04/02/2008.

[11] Decreto Protezione Civile Regione Sicilia n.3105 del 7/02/2001 G.U. n.43 del 5/03 2001

[12] Caruso F, Pluchino A, Latora V, Vinciguerra S, Rapisarda A. A new analysis of Self-Organized Criticality in the OFC model and in real earthquakes. *Phys Rev E* **75** (2007) 055101(R)

[13] Barbarossa L, Privitera R, Martinico F. Insediamenti irregolari e rischi territoriali lungo i litorali del Val di Noto. Percorsi di progetto per la città costiera resiliente. Atti della XIX Conferenza Siu Cambiamenti, Catania 16 -18 giugno 2016. Planum Publisher, 2016